\documentstyle[12pt]{article}
\def\lb{\label}
\def\be{\begin{equation}}
\def\ee{\end{equation}}
\newcommand{\bin}[2]{
\begin{array}{c}
#1 \\ #2
\end{array}}
\begin{document}


\title{Tetrahedron Reflection Equation.
\thanks{Preprint LPTHE - 96 - 41}
}

\author{A.P. Isaev
\\
{\it Bogoliubov Theoretical Laboratory, JINR, Dubna,} \\
{\it 141 980, Moscow Region, Russia} \\
P.P.Kulish \thanks{\it On leave of absence
>from the Steklov Mathematical Institute,
Fontanka 27, St.Petersburg, 191011, Russia} \\
{\it LPTHE, Universite P. et M. Curie, URA 280, C.N.R.S.,}\\
{\it Tour 16, 4, pl.Jussieu, F-75252, Paris, Cedex 05, France}
}

\date{}

\maketitle

\begin{abstract}
Reflection equation for the scattering of lines moving in half-plane
is obtained. The corresponding geometric picture is related with
configurations of half-planes touching the boundary plane in 2+1
dimensions. This equation can be obtained as an additional to the
tetrahedron equation consistency condition for a modified Zamolodchikov
algebra.
\end{abstract}


\section{Introduction}

Active recent development of the reflection equation in 1+1 dimension
gives rise to variety of interesting physical results on solvable models
with non-periodic boundary conditions
and mathematical relations such as braid group in a solid handlebody
(see \cite{PPK} and Refs therein). Having this in mind and the existing
of restricted, although quite elaborated, results
(see \cite{SMS} - \cite{LM}) for the Zamolodchikov
tetrahedron equation (TE) \cite{Zam},
a straightforward generalization
of the reflection equation for the 2+1 dimension (tetrahedron reflection
equation) is given in this paper.
Moving to higher dimensions we follow the kinematic analogy
transforming the objects corresponding to events in
smaller dimensions to
generators of a Zamolodchikov algebra while defining
relations for them acquire an extra exchange algebra factor; e.g.
>from 1 dim to 2 dim we have:
$$
A \cdot A  \equiv A \cdot A \, \rightarrow  \,
A_1 \, A_2 = R_{12} \, A_2 \, A_1 \; .
$$
So the commutativity condition is transformed into the Zamolodchikov
(exchange) algebra. In the same manner,
the Yang-Baxter equation for 2 dim is transformed
into the Zamolodchikov algebra for moving straight lines on a plane
($2+1$ dimensions):
$$
A^{(12)} \, A^{(13)} \, A^{(23)} =
A^{(23)} \, A^{(13)} \, A^{(12)} \, \rightarrow  \,
A^{(12)}_{1} \, A^{(13)}_{2} \, A^{(23)}_{3} =
R_{123} \, A_{3}^{(23)} \, A_{2}^{(13)} \, A_{1}^{(12)} \; ,
$$
giving rise to the tetrahedron equation
as a consistency condition. This procedure can be
extended for higher dimensions \cite{BaSt}.
We note that 2-dim reflection equation
(without spectral parameters) has the form of commutativity
condition
$$
K_{1} \cdot K_{12} = K_{12} \cdot K_{1} \; ,
$$
where $K_{12} = R_{21} \, K_{2} \, R_{12}$ are dressed
reflection matrices. It is remarkable that reflection
equation for $2+1$ dimensions has the form of the Yang-Baxter
equation but for dressed operators too (see Sect. 3).

As we will see below the geometrical picture is also
helpful for the construction, where
trajectories from smaller dimension started to be moving objects in higher
dimension.

\section{Zamolodchikov Algebra with Boundary Operators
in 3 Dimensions}

We recall \cite{GZ} that factorizable scattering (in two-dimensional
integrable quantum field theory with boundary conditions)
can be described by the Zamolodchikov algebra with
generators $\{ A_{i}(x) \}$ and boundary operator $B$
which satisfy the defining relations
\be
\lb{ikk}
A_{i}(x) \, A_{j}(y) = R_{ij}^{kl}(x,y) \, A_{l}(y) \, A_{k}(x) \; , \;\;
A_{i}(x) \, B = K_{i}^{j}(x) \, A_{j}(\overline{x}) \, B \; .
\ee
The consistency conditions for this algebra give rise to
the Yang-Baxter equations for matrices $R$
$$
R^{i_{1}i_{2}}_{j_{1}j_{2}}(x,y) \,
R^{j_{1}i_{3}}_{k_{1}j_{3}}(x,z) \,
R^{j_{2}j_{3}}_{k_{2}k_{3}}(y,z) =
R^{i_{2}i_{3}}_{j_{2}j_{3}}(y,z) \,
R^{i_{1}j_{3}}_{j_{1}k_{3}}(x,z) \,
R^{j_{1}j_{2}}_{k_{1}k_{2}}(x,y)  \Rightarrow
$$
$$
R_{12}(x,y) \, R_{13}(x,z) \, R_{23}(y,z)  =
R_{23}(y,z) \, R_{13}(x,z) \, R_{12}(x,y) \; ,
$$
and reflection equations for
$R$ and $K$
\be
\lb{2dRE}
K_{2}(y) \, R_{12}(x,\overline{y}) \,
K_{1}(x) \, R_{21}(\overline{y},\overline{x})  =
R_{12}(x,y) \, K_{1}(x) \, R_{21}(y,\overline{x}) \, K_{2}(y) \; .
\ee
Here and below we use standard matrix notations
of the $R$-matrix formalism \cite{FRT}.
Our aim, in this section, is to define 3d
analogues of the algebra (\ref{ikk}).

Let us consider the picture:

\unitlength=6mm
\begin{picture}(25,9)

\put(4,7.5){$V_{III} \rightarrow$}
\put(8,7.5){$V_{II} \rightarrow$}
\put(10.5,7.5){$V_{I} \rightarrow$}
\put(4,6){$\theta_{III}$}
\put(8.2,6){$\theta_{II}$}
\put(11,6){$\theta_{I}$}

\put(0,7){\line(1,0){15}}

\put(4,7){\line(-2,-1){4}}
\put(4,7){\vector(2,-1){12}}

\put(0,5.5){$\overline{z}$}
\put(2,2){$\overline{y}$}

\put(8.5,7){\line(-1,-1){6}}
\put(8.5,7){\vector(1,-1){6}}

\put(8,1){$\overline{x}$}

\put(10.5,7){\line(-1,-4){1.7}}
\put(10.5,7){\vector(1,-4){1.7}}

\put(11.2,0.5){$x$}
\put(13.8,0.5){$y$}
\put(15.5,1.5){$z$}

\put(6,5.1){$1$}
\put(9,3.5){$2$}
\put(9.2,5.2){$3$}
\put(11.5,4.2){$4$}
\put(11,2.5){$5$}
\put(13.5,2.5){$6$}

\put(18,0.5){\bf Fig. 1}

\end{picture}

We have 3 broken lines (b-lines) with peaks $III,II,I$
moving from left to right with velocities
$V_{III} > V_{II} > V_{I}=0$. For convenience we
will think that each b-line has
2 colours, namely segments of b-line $I$ have colours $(x, \; \overline{x})$,
b-line $II$ has colours $(y, \; \overline{y})$ and
b-line $III$ has colours $(z, \; \overline{z})$. We denote
angles between segments of b-lines as
$\theta_{III,II,I}$ ($\theta_{III} > \theta_{II} > \theta_{I}$).
One can consider colours $x, \, y, \, z$ as two-vectors
(velocities of falling strings;
$\overline{x}$, $\overline{y}$ and $\overline{z}$ are
velocities of reflecting strings)
which yield another equivalent parametrization of b-lines
instead of $\{ V_{I}, \, \theta_{I} \}$,
$\{ V_{II}, \, \theta_{II} \}$ and $\{ V_{III}, \, \theta_{III} \}$.
Numbers $1,2, \dots,6$ denote intersections of the b-line segments.
It is clear that one can relate these numbers with pairs of colours:
\be
\lb{ik.1}
(y, \, z)=6 \; , \;\;
(x, \, z)=5 \; , \;\;
(x, \, y)=4 \; , \;\;
(\overline{x}, \, y)=3 \; , \;\;
(\overline{x}, \, z)=2 \; , \;\;
(\overline{y}, \, z)=1 \; .
\ee
We also ascribe vector index
to each intersection and, therefore, $(1,2, \dots,6)$
can be considered as the numbers of vector spaces.

The following events can happen
when b-lines move from left to right

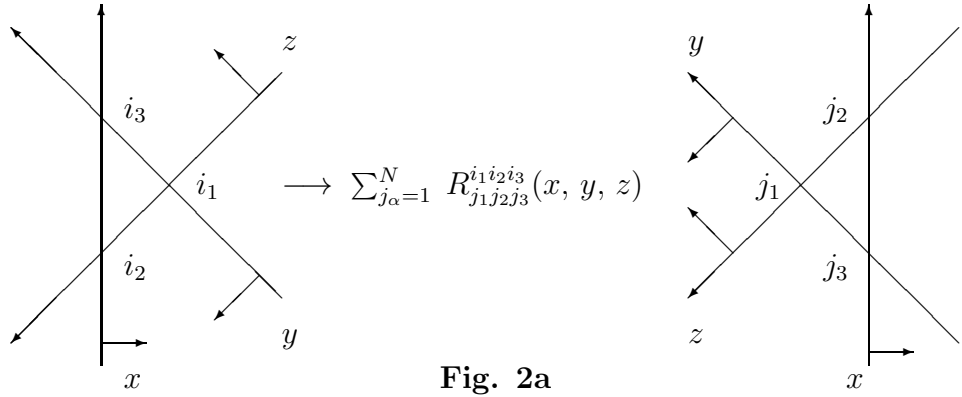
\unitlength=6mm
\begin{picture}(25,10)

\put(3,1.5){\vector(1,0){1}}
\put(6.5,3){\vector(-1,-1){1}}
\put(6.5,7){\vector(-1,1){1}}

\put(3,1){\vector(0,1){8}}
\put(1,8.5){\line(1,-1){6}}
\put(2,7.5){\vector(-1,1){1}}
\put(1,1.5){\line(1,1){6}}
\put(2,2.5){\vector(-1,-1){1}}

\put(3.5,0.5){$x$}
\put(7,1.5){$y$}
\put(7,8){$z$}

\put(3.5,3){$i_2$}
\put(3.5,6.5){$i_3$}
\put(5.1,4.8){$i_1$}

\put(7,4.8){$ \longrightarrow \; \sum_{j_{\alpha}=1}^{N}
\; R^{i_{1}i_{2}i_{3}}_{j_{1}j_{2}j_{3}}(x, \, y, \, z) $}

\put(20,1.3){\vector(1,0){1}}
\put(17,6.5){\vector(-1,-1){1}}
\put(17,3.5){\vector(-1,1){1}}

\put(20,1){\vector(0,1){8}}
\put(22,8.5){\vector(-1,-1){6}}
\put(22,1.5){\vector(-1,1){6}}

\put(19.5,0.5){$x$}
\put(16,8){$y$}
\put(16,1.5){$z$}

\put(19,3){$j_3$}
\put(19,6.5){$j_2$}
\put(17.5,4.8){$j_1$}

\put(10.5,0.5){\bf Fig. 2a}

\end{picture}

\unitlength=6mm
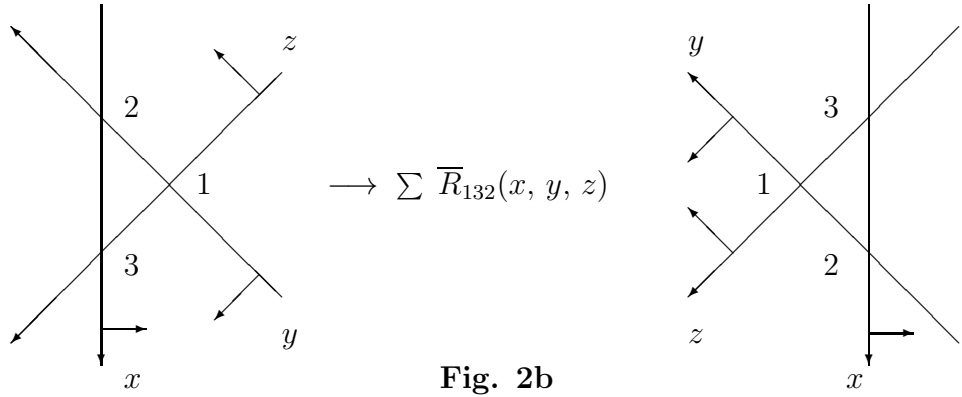
\begin{picture}(25,10)

\put(3,1.8){\vector(1,0){1}}
\put(6.5,3){\vector(-1,-1){1}}
\put(6.5,7){\vector(-1,1){1}}

\put(3,9){\vector(0,-1){8}}
\put(1,8.5){\line(1,-1){6}}
\put(2,7.5){\vector(-1,1){1}}
\put(1,1.5){\line(1,1){6}}
\put(2,2.5){\vector(-1,-1){1}}

\put(3.5,0.5){$x$}
\put(7,1.5){$y$}
\put(7,8){$z$}

\put(3.5,3){$3$}
\put(3.5,6.5){$2$}
\put(5.1,4.8){$1$}

\put(8,4.8){$ \longrightarrow \; \sum \;
\overline{R}_{132}(x, \, y, \, z) $}

\put(20,1.7){\vector(1,0){1}}
\put(17,6.5){\vector(-1,-1){1}}
\put(17,3.5){\vector(-1,1){1}}

\put(20,9){\vector(0,-1){8}}
\put(22,8.5){\vector(-1,-1){6}}
\put(22,1.5){\vector(-1,1){6}}

\put(19.5,0.5){$x$}
\put(16,8){$y$}
\put(16,1.5){$z$}

\put(19,3){$2$}
\put(19,6.5){$3$}
\put(17.5,4.8){$1$}

\put(10.5,0.5){\bf Fig. 2b}

\end{picture}

\noindent
and also (when $V_{II} > V_{I}$) we have

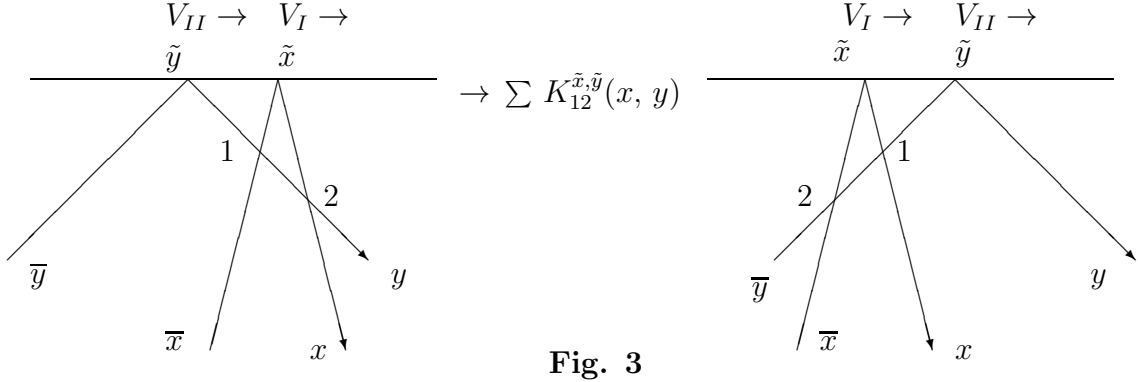
\unitlength=6mm
\begin{picture}(25,10)

\put(2,8.2){$V_{II} \rightarrow$}
\put(4.5,8.2){$V_{I} \rightarrow$}
\put(2,7.3){$\tilde{y}$}
\put(4.5,7.3){$\tilde{x}$}

\put(-1,7){\line(1,0){9}}

\put(-1,2.5){$\overline{y}$}

\put(2.5,7){\line(-1,-1){4}}
\put(2.5,7){\vector(1,-1){4}}

\put(2,1){$\overline{x}$}

\put(4.5,7){\line(-1,-4){1.5}}
\put(4.5,7){\vector(1,-4){1.5}}

\put(5.2,0.8){$x$}
\put(7,2.5){$y$}

\put(3.2,5.2){$1$}
\put(5.5,4.2){$2$}

\put(8.5,6.5){$\rightarrow \, \sum \,
K_{12}^{\tilde{x},\tilde{y}}(x, \, y)$}

\put(17,8.2){$V_{I} \rightarrow$}
\put(19.5,8.2){$V_{II} \rightarrow$}
\put(16.8,7.4){$\tilde{x}$}
\put(19.5,7.4){$\tilde{y}$}

\put(14,7){\line(1,0){9}}

\put(15,2.2){$\overline{y}$}

\put(19.5,7){\line(-1,-1){4}}
\put(19.5,7){\vector(1,-1){4}}

\put(16.5,1){$\overline{x}$}

\put(17.5,7){\line(-1,-4){1.5}}
\put(17.5,7){\vector(1,-4){1.5}}

\put(19.5,0.8){$x$}
\put(22.5,2.5){$y$}

\put(18.2,5.2){$1$}
\put(16,4.2){$2$}

\put(10.5,0.5){\bf Fig. 3}

\end{picture}

\noindent
Here $K_{12}^{\tilde{x},\tilde{y}}(x, \, y)$ is an operator
which characterizes the reflecting of pair of strings from the border,
indices $1,2$ denote matrix spaces, $x,y$ are two-vectors
(vector velocities of falling segments of
broken strings) and $\tilde{x},\tilde{y}$ - numbers of
operator spaces.

One can rewrite Figs. 2a, 2b in the algebraic form of defining
relations for 3d Zamolodchikov algebra (see e.g. \cite{Kor})
\be
\lb{ik.1a}
A_{1}^{\hat{z},\hat{y}}(z,y) \,
A_{2}^{\hat{z},\hat{x}}(z,x) \,
A_{3}^{\hat{y},\hat{x}}(y,x) =
R_{123}(x,y,z) \,
A_{3}^{\hat{y},\hat{x}}(y,x) \,
A_{2}^{\hat{z},\hat{x}}(z,x) \,
A_{1}^{\hat{z},\hat{y}}(z,y) \; ,
\ee
\be
\lb{ik.1b}
A_{1}^{\hat{z},\hat{y}}(z,y) \,
A_{2}^{\hat{x},\hat{y}}(x,y) \,
A_{3}^{\hat{x},\hat{z}}(x,z) =
\overline{R}_{132}(x,y,z) \,
A_{3}^{\hat{x},\hat{z}}(x,z) \,
A_{2}^{\hat{x},\hat{y}}(x,y) \,
A_{1}^{\hat{z},\hat{y}}(z,y) \; ,
\ee

where $N$- vectors (in $i$-th matrix space)

\begin{picture}(20,5)

\put(2,2){$A_{i}^{\hat{x},\hat{y}}(x,y) \equiv $}

\put(10.5,3){$y$}
\put(7,3){$x$}

\put(10,3){\vector(-1,-1){2}}
\put(8,3){\vector(1,-1){2}}

\put(8.2,1.8){$i$}
\end{picture}

\noindent
are generators of the 3d Zamolodchikov
algebra, $x,y$ are 2d vector- velocities, while
indices $\hat{x},\hat{y}$ denote the numbers of
auxiliary operator spaces,
and we omit these indices in formulas below
in view of their one to one correspondence
with velocities.
It is clear that the kind of "unitarity" condition is valid
\be
\lb{RR}
\overline{R}_{132}(x, \, y, \, z) =
R^{-1}_{321}(y, \, z, \, x) \; ,
\ee
and the matrices
$R_{123}(x, \, y, \, z) \in (Mat(N))^{\otimes 3}$
are solutions of the tetrahedron equation
\be
\lb{ik.2}
\begin{array}{c}
R_{123}(x,y,z) \, R_{145}(u,y,z) \,
R_{246}(u,x,z) \, R_{356}(u,x,y) = \\ \\
R_{356}(u,x,y) \, R_{246}(u,x,z) \,
R_{145}(u,y,z) \, R_{123}(x,y,z) \; ,
\end{array}
\ee
which can be derived (see \cite{Zam})
if one consider two possible ways
(depending on what is the first - to turn out inside
the triangle $A$ or $B$)
to transform the picture

\unitlength=6mm
\begin{picture}(25,10)

\put(2,9.5){\vector(1,-3){3}}
\put(5,9.5){\vector(-1,-3){3}}
\put(1,8){\vector(2,-1){7}}
\put(1,2){\line(2,1){7}}
\put(2,2.5){\vector(-2,-1){1}}

\put(5.5,9){$x$}
\put(1.5,9){$y$}
\put(0.5,1.5){$u$}
\put(0.5,8){$z$}

\put(4.5,3){$5$}
\put(4.5,6.5){$2$}
\put(3,4.8){$3$}
\put(6,4.8){$4$}
\put(2.8,2){$6$}
\put(2.8,7.5){$1$}
\put(3.2,3.5){$B$}
\put(3.2,6){$A$}

\put(11,4.8){$ \; \longrightarrow \; $}

\put(18,9.5){\line(1,-3){3}}
\put(21,9.5){\line(-1,-3){3}}
\put(22,8){\line(-2,-1){7}}
\put(22,2){\line(-2,1){7}}

\put(20,9){$x$}
\put(17,9){$y$}
\put(15,6){$z$}
\put(15,3.5){$u$}

\put(18.2,3){$2$}
\put(18.2,6.5){$5$}
\put(19.8,4.8){$3$}
\put(16.8,4.8){$4$}
\put(20.5,2.8){$1$}
\put(20.5,6.7){$6$}

\put(10.5,0.5){\bf Fig. 4}

\end{picture}

\noindent
or reorder corresponding monomial of sixth degree of the
3d Zamolodchikov algebra
$$
A_{1}(z,y) \, A_{2}(z,x) \, A_{3}(y,x) \, A_{4}(z,u) \,
A_{5}(y,u) \, A_{6}(x,u) \; .
$$
Note, that the tetrahedron eq. (\ref{ik.2}) has
simple solutions
$R_{123}(x,y,z) = I_{1} \, R_{23}(y,z)$,
$R_{123}(x,y,z) = R_{12}(x,y) \, I_{3}$, where $R_{ij}(x,y)$
is arbitrary solution of the Yang-Baxter equation.
In fact, one can rewrite (\ref{ik.2}) in the following
concise form
\be
\lb{ik.3}
\begin{array}{c}
R(x,y,z) \, R(u,y,z) \,
R(u,x,z) \, R(u,x,y) = \\ \\
R(u,x,y) \, R(u,x,z) \,
R(u,y,z) \, R(x,y,z) \; ,
\end{array}
\ee
since the matrix indices $1,2,\dots,6$ for $R$'s can be restored
uniquely from the order of arguments $x,y,z,u$. Another
convenient form is
$$
\begin{array}{c}
R_{123}(x,y,z) \, \left[ \hat{R}_{1}'(u,y,z) \,
\hat{R}_{2}(u,x,z) \, \hat{R}_{3}'(u,x,y) \right] = \\ \\
\left[ \hat{R}_{3}(u,x,y) \, \hat{R}_{2}'(u,x,z) \,
\hat{R}_{1}(u,y,z) \right] \, R_{123}(x,y,z) \; ,
\end{array}
$$
where $\hat{R}_{i}' = {\cal P}_{56} \, R_{i56}$,
$\hat{R}_{i} = {\cal P}_{45} \, R_{i45}$ and
${\cal P}_{12}$ is the permutation matrix.

One can also represent Fig. 3 in algebraical form with
the help of boundary operators $B^{\tilde{x}}(x)$.
Namely we have for Fig. 3 the representation
\be
\lb{ik.33}
B^{\tilde{y}}(y) \, A_{1}(\overline{x},y) \,
B^{\tilde{x}}(x) \, A_{2}(y,x) = K^{\tilde{x}\tilde{y}}_{12}(x,y) \,
A_{2}(\overline{x},\overline{y}) \,
B^{\tilde{x}}(x) \, A_{1}(\overline{y},x) \, B^{\tilde{y}}(y) \; ,
\ee
which generalizes 2d reflection equation \cite{KS}
(we obtain 2d reflection equation (\ref{2dRE})
if we put $K_{12} =1$ in (\ref{ik.33})).
We stress that
we try to conserve the clockwise rule in writing indices
in all algebraical analogues of the pictures given above.

At the end of this discussion we present $RTTT$ relations
which are 3d analogues of $RTT$ relations
(about $RTT$ relations see \cite{FRT}):
\be
\lb{ik.11a}
\begin{array}{c}
R_{123}(x,y,z) \,
T_{1}^{\hat{z},\hat{y}}(z,y) \,
T_{2}^{\hat{z},\hat{x}}(z,x) \,
T_{3}^{\hat{y},\hat{x}}(y,x) \, = \\
= T_{3}^{\hat{y},\hat{x}}(y,x) \,
T_{2}^{\hat{z},\hat{x}}(z,x) \,
T_{1}^{\hat{z},\hat{y}}(z,y) \,
R_{123}(x,y,z) \; .
\end{array}
\ee
Here $T_{i}(.) \in Mat_{i}(N)$ is the matrix in $i$-th
matrix space. Comparing (\ref{ik.2}) with (\ref{ik.11a})
one can find the matrix representations for $T_{i}$ operators
$$
T^{\hat{z}\hat{y}}_{1}(z,y) = R_{1\hat{z}\hat{y}}(u,y,z) \; , \;\;
T^{\hat{z}\hat{y}}_{1}(z,y) = R^{-1}_{\hat{z}\hat{y}1}(y,z,u) =
\overline{R}_{1\hat{z}\hat{y}}(u,y,z) \; .
$$

In this section, we have obtained the algebra with
generators $A_{1}^{\hat{x}\hat{y}}(x,y)$
and $B^{\tilde{x}}(x)$ and defining relations
(\ref{ik.1a}), (\ref{ik.33}).
This algebra is 3d analogue
of the Zamolodchikov algebra with boundary operator $B$ (\ref{ikk}).
In the next section, we show that the consistency
conditions for the algebra
(\ref{ik.1a}), (\ref{ik.33})
include not only tetrahedron equation
(\ref{ik.2}) but also new relation for the scattering
and reflection matrices $R$ and $K$ which can be
interpreted as 3 dimensional reflection equation.

\section{3d Reflection Equations}

As one can see (e.g. from considering of Fig. 1) there are
different scenarios of strings reflecting from the border.
Namely, after the reordering of velocities $V_{III} > V_{II} > V_{I}$
we have 6 different choices of angles
$\theta_{III} > \theta_{II} > \theta_{I}$,
$\theta_{II} > \theta_{III} > \theta_{I}$ etc. It is not obvious
that all of these choices give the relations which can be
sewed in one 3d reflection equation. Nevertheless, below
we show that all of these 6 scenarios define the
unique tetrahedron reflection equation.

\vspace{0.5cm}

{\bf Scenario 1.} \\

Here we consider two possible ways (depending on initial data)
of moving b-lines on the picture Fig. 1.
Namely the first way is when the top of b-line $III$
reaches the top of b-line $II$, then reaches the top of b-line $I$ and only
then the top of b-line $II$ interacts with the top of b-line $I$.
Another way is when we have the sequence $(II,I), \; (III,I), \;
(III,II)$. The algebraic expression related to Fig. 1
(directions of arrows on strings are important here) has the form
\be
\lb{ik.33a}
1) \;\; B^{\tilde{z}}(z) \, A_{1}(\overline{y},z) \,
B^{\tilde{y}}(y) \, A_{2}(\overline{x},z) \,
A_{3}(\overline{x},y) \, B^{\tilde{x}}(x) \,
A_{4}(y,x) \, A_{5}(z,x) \, A_{6}(z,y) \,
\; ,
\ee
Reordering this monomial to the new form
\be
\lb{ik.33b}
A_{6}(\overline{y},\overline{z})  \,
A_{5}(\overline{x},\overline{z}) \,
A_{4}(\overline{x},\overline{y}) \,
B^{\tilde{x}}(x) \, A_{3}(\overline{y},x) \,
A_{2}(\overline{z},x) \, B^{\tilde{y}}(y) \,
A_{1}(\overline{z},y) \, B^{\tilde{z}}(z) \,
\ee
(accordingly
with 2 different ways described above)
with the help of the rules presented on Figs. 2a, 2b, 3
and taking into account that the result should be
independent of these ways (of initial data) we obtain the equation
($\theta_{III} > \theta_{II} > \theta_{I}$)
\be
\lb{ik.4}
\begin{array}{c}
\overline{R}_{465}(z,x,y) \, R_{236}(y,z,\overline{x}) \,
K_{16}^{\tilde{y}\tilde{z}}(y,z) \,
K_{25}^{\tilde{x}\tilde{z}}(x,z) \,
\overline{R}_{153}(\overline{x},y,\overline{z}) \,
R_{124}(x,y,\overline{z}) \,
K_{34}^{\tilde{x}\tilde{y}}(x,y) =
\\ \\
K_{34}^{\tilde{x}\tilde{y}}(x,y) \,
\overline{R}_{142}(\overline{x},z,\overline{y}) \,
R_{135}(x,z,\overline{y}) \,
K_{25}^{\tilde{x}\tilde{z}}(x,z) \,
K_{16}^{\tilde{y}\tilde{z}}(y,z) \,
\overline{R}_{263}(\overline{y},x,\overline{z}) \,
R_{456}(\overline{z},\overline{y},\overline{x}) \; .
\end{array}
\ee

\vspace{0.5cm}

{\bf Scenario 2.} \\

Now we take the case $\theta_{y} > \theta_{z} > \theta_{x}$
(here and below we use indices $x,y,z$ instead of $I,II,III$).
The related picture of the type Fig. 1
leads to the consideration of the monomial
$$
2) \;\; A_{1}(\overline{z},\overline{y}) \, B^{\tilde{z}}(z) \,
A_{2}(\overline{y},z) \,
B^{\tilde{y}}(y) \, A_{3}(\overline{x},z) \,
A_{4}(\overline{x},y) \, B^{\tilde{x}}(x) \,
A_{5}(y,x) \, A_{6}(z,x) \; ,
$$
which is necessary to reorder. For this we take into
account relations Figs. 2a, 2b, 3 and also new relation ($V_{y} > V_{x}$)

\unitlength=6mm
\begin{picture}(25,10)

\put(2,8.2){$V_{y} \rightarrow$}
\put(4.5,8.2){$V_{x} \rightarrow$}
\put(1.8,7.4){$\tilde{y}$}
\put(4.5,7.4){$\tilde{x}$}

\put(-1,7){\line(1,0){9}}

\put(0,2.2){$\overline{x}$}

\put(4.5,7){\line(-1,-1){4}}
\put(4.5,7){\vector(1,-1){4}}

\put(1.5,1){$\overline{y}$}

\put(2.5,7){\line(-1,-4){1.5}}
\put(2.5,7){\vector(1,-4){1.5}}

\put(4.5,0.8){$y$}
\put(7.5,2.5){$x$}

\put(3.2,5.2){$2$}
\put(1,4.2){$1$}

\put(8.5,6.5){$\rightarrow \, \sum \,
\overline{K}_{12}^{\tilde{x},\tilde{y}}(x, \, y)$}

\put(17,8.2){$V_{x} \rightarrow$}
\put(19.5,8.2){$V_{y} \rightarrow$}
\put(17,7.3){$\tilde{x}$}
\put(19.5,7.3){$\tilde{y}$}

\put(14,7){\line(1,0){9}}

\put(14,2.5){$\overline{x}$}

\put(17.5,7){\line(-1,-1){4}}
\put(17.5,7){\vector(1,-1){4}}

\put(17,1){$\overline{y}$}

\put(19.5,7){\line(-1,-4){1.5}}
\put(19.5,7){\vector(1,-4){1.5}}

\put(20.2,0.8){$y$}
\put(22,2.5){$x$}

\put(18.2,5.2){$2$}
\put(20.5,4.2){$1$}

\put(10.5,0.5){\bf Fig. 5}

\end{picture}

\noindent
which is equivalent to the algebraic formula
\be
\lb{ik.55}
A_{1}(\overline{y},\overline{x}) \,
B^{\tilde{y}}(y) \, A_{2}(\overline{x},y) \,
B^{\tilde{x}}(x) \,
= \overline{K}_{12}^{\tilde{x}\tilde{y}}(x,y) \,
B^{\tilde{x}}(x) \, A_{2}(\overline{y},x) \,
B^{\tilde{y}}(y) \, A_{1}(x,y) \; .
\ee
It is evident (from comparing
(\ref{ik.33}) and (\ref{ik.55}))
that the kind of unitarity condition holds
\be
\lb{KK}
\overline{K}_{12}(x,y) = (K_{21}(y,x))^{-1} \; .
\ee
Now we obtain the following reflection equation
\be
\lb{ik.5}
\begin{array}{c}
K_{45}^{\tilde{x}\tilde{y}}(x,y) \,
\overline{R}_{253}(\overline{x},z,\overline{y}) \,
R_{246}(x,z,\overline{y}) \,
K_{36}^{\tilde{x}\tilde{z}}(x,z) \,
\overline{R}_{165}(\overline{x},\overline{y},\overline{z}) \,
R_{134}(x,\overline{y},\overline{z}) \,
\overline{K}_{12}^{\tilde{y}\tilde{z}}(y,z) =
\\ \\
\overline{K}_{12}^{\tilde{y}\tilde{z}}(y,z) \,
\overline{R}_{143}(\overline{x},z,y) \,
R_{156}(x,z,y) \,
K_{36}^{\tilde{x}\tilde{z}}(x,z) \,
\overline{R}_{264}(\overline{x},y,\overline{z}) \,
R_{235}(x,y,\overline{z}) \,
K_{45}^{\tilde{x}\tilde{y}}(x,y) \,
\; .
\end{array}
\ee


{\bf Scenarios 3,4,5,6.} \\

In the same way one can consider
the cases 3.) $\theta_{x} > \theta_{y} > \theta_{z}$,
4.) $\theta_{y} > \theta_{x} > \theta_{z}$,
5.) $\theta_{x} > \theta_{z} > \theta_{y}$ and
6.) $\theta_{z} > \theta_{x} > \theta_{y}$.
The corresponding monomials which are necessary to reorder
can be written as
$$
3) \;\;
A_{1}(\overline{z},\overline{y}) \, A_{2}(\overline{z},\overline{x}) \,
B^{\tilde{z}}(z) \, A_{3}(\overline{x},z) \, A_{4}(\overline{y},z) \,
A_{5}(\overline{y},\overline{x}) \,
B^{\tilde{y}}(y) \, A_{6}(\overline{x},y) \, B^{\tilde{x}}(x) \; ,
$$
$$
4) \;\;
A_{1}(\overline{z},\overline{x}) \, A_{2}(\overline{z},\overline{y}) \,
B^{\tilde{z}}(z) \, A_{3}(\overline{y},z) \, A_{4}(\overline{x},z) \,
B^{\tilde{y}}(y) \, A_{5}(\overline{x},y) \,
B^{\tilde{x}}(x) \, A_{6}(y,x) \; ,
$$
$$
5) \;\;
A_{1}(\overline{z},\overline{x}) \, B^{\tilde{z}}(z) \,
A_{2}(\overline{x},z) \,
A_{3}(\overline{y},z) \, A_{4}(\overline{y},\overline{x}) \,
B^{\tilde{y}}(y) \, A_{5}(\overline{x},y) \, A_{6}(z,y) \,
B^{\tilde{x}}(x) \; ,
$$
$$
6) \;\;
B^{\tilde{z}}(z) \, A_{1}(\overline{x},z) \,
A_{2}(\overline{y},z) \,
A_{3}(\overline{y},\overline{x}) \, B^{\tilde{y}}(y) \,
A_{4}(\overline{x},y) \, A_{5}(z,y) \,
B^{\tilde{x}}(x) \, A_{6}(z,x) \,
\; .
$$
The reordering of these monomials in two possible ways
gives us the corresponding 3d reflection equations
\be
\lb{ik.6}
\begin{array}{c}
\overline{R}_{354}(\overline{y},z,\overline{x}) \,
R_{125}(\overline{x},\overline{y},\overline{z}) \,
\overline{K}_{14}^{\tilde{y}\tilde{z}}(y,z) \,
\overline{R}_{163}(\overline{x},z,y) \,
R_{246}(y,\overline{x},\overline{z}) \,
\overline{K}_{23}^{\tilde{x}\tilde{z}}(x,z) \,
\overline{K}_{56}^{\tilde{x}\tilde{y}}(x,y) =
\\ \\
\overline{K}_{56}^{\tilde{x}\tilde{y}}(x,y) \,
\overline{K}_{23}^{\tilde{x}\tilde{z}}(x,z) \,
\overline{R}_{264}(\overline{y},z,x) \,
R_{136}(x,\overline{y},\overline{z}) \,
\overline{K}_{14}^{\tilde{y}\tilde{z}}(y,z) \,
\overline{R}_{152}(x,z,y) \,
R_{345}(y,x,\overline{z}) \; .
\end{array}
\ee

\be
\lb{ik.7}
\begin{array}{c}
\overline{K}_{23}^{\tilde{y}\tilde{z}}(y,z) \,
\overline{R}_{254}(\overline{x},z,y) \,
R_{135}(y,\overline{x},\overline{z}) \,
\overline{K}_{14}^{\tilde{x}\tilde{z}}(x,z) \,
\overline{R}_{162}(y,z,x) \,
R_{346}(x,y,\overline{z}) \,
K_{56}^{\tilde{x}\tilde{y}}(x,y) =
\\ \\
K_{56}^{\tilde{x}\tilde{y}}(x,y) \,
\overline{R}_{364}(\overline{x},z,\overline{y}) \,
R_{126}(\overline{y},\overline{x},\overline{z}) \,
\overline{K}_{14}^{\tilde{x}\tilde{z}}(x,z) \,
\overline{R}_{153}(\overline{y},z,x) \,
R_{245}(x,\overline{y},\overline{z}) \,
\overline{K}_{23}^{\tilde{y}\tilde{z}}(y,z) \; .
\end{array}
\ee

\be
\lb{ik.8}
\begin{array}{c}
\overline{R}_{243}(\overline{y},z,\overline{x}) \,
R_{256}(y,z,\overline{x}) \,
K_{36}^{\tilde{y}\tilde{z}}(y,z) \,
\overline{R}_{164}(\overline{y},\overline{x},\overline{z}) \,
R_{135}(y,\overline{x},\overline{z}) \,
\overline{K}_{12}^{\tilde{x}\tilde{z}}(x,z) \,
\overline{K}_{45}^{\tilde{x}\tilde{y}}(x,y) =
\\ \\
\overline{K}_{45}^{\tilde{x}\tilde{y}}(x,y) \,
\overline{K}_{12}^{\tilde{x}\tilde{z}}(x,z) \,
\overline{R}_{153}(\overline{y},z,x) \,
R_{146}(y,z,x) \,
K_{36}^{\tilde{y}\tilde{z}}(y,z) \,
\overline{R}_{265}(\overline{y},x,\overline{z}) \,
R_{234}(y,x,\overline{z}) \; .
\end{array}
\ee

\be
\lb{ik.9}
\begin{array}{c}
\overline{K}_{34}^{\tilde{x}\tilde{y}}(x,y) \,
\overline{R}_{365}(z,y,x) \,
R_{246}(x,z,\overline{y}) \,
K_{16}^{\tilde{x}\tilde{z}}(x,z) \,
K_{25}^{\tilde{y}\tilde{z}}(y,z) \,
\overline{R}_{154}(\overline{y},x,\overline{z}) \,
R_{123}(y,x,\overline{z}) =
\\ \\
\overline{R}_{132}(\overline{y},z,\overline{x}) \,
R_{145}(y,z,\overline{x}) \,
K_{25}^{\tilde{y}\tilde{z}}(y,z) \,
K_{16}^{\tilde{x}\tilde{z}}(x,z) \,
\overline{R}_{264}(\overline{x},y,\overline{z}) \,
R_{356}(\overline{z},\overline{x},\overline{y}) \,
\overline{K}_{34}^{\tilde{x}\tilde{y}}(x,y) \; .
\end{array}
\ee

Now one can substitute relations (\ref{RR}) and (\ref{KK})
into eqs. (\ref{ik.4}), (\ref{ik.5})-(\ref{ik.9}). Then we see
that all these equations (written in different forms)
are identical after appropriate permutation
of indices $1, \dots, 6$ and velocities $x,y,z$.
Therefore, one can explore only one of them.
We take eq. (\ref{ik.4}) and substitute
there (\ref{RR}) and definition
$$
K_{12}^{\tilde{x}\tilde{y}}(x,y) \equiv
{\cal P}_{12} \, \hat{K}_{12}^{\tilde{x}\tilde{y}}(x,y) \; .
$$
As a result we obtain the following
form of 3d reflection equation:
\be
\lb{ik.10}
\begin{array}{c}
R^{-1}_{123}(x,y,z) \, R_{541}(y,z,\overline{x}) \,
\hat{K}_{16}^{\tilde{y}\tilde{z}}(y,z) \,
\hat{K}_{25}^{\tilde{x}\tilde{z}}(x,z) \,
R^{-1}_{541}(y,\overline{z},\overline{x}) \,
R_{123}(x,y,\overline{z}) \,
\hat{K}_{34}^{\tilde{x}\tilde{y}}(x,y) =
\\ \\
\hat{K}_{34}^{\tilde{x}\tilde{y}}(x,y) \,
R^{-1}_{456}(z,\overline{y},\overline{x}) \,
R_{632}(x,z,\overline{y}) \,
\hat{K}_{25}^{\tilde{x}\tilde{z}}(x,z) \,
\hat{K}_{16}^{\tilde{y}\tilde{z}}(y,z) \,
R^{-1}_{632}(x,\overline{z},\overline{y}) \,
R_{456}(\overline{z},\overline{y},\overline{x}) \; .
\end{array}
\ee
Let us put the condition
$R_{123}(x,y,z) =
R_{123}(\overline{x},\overline{y},\overline{z})$,
which is equivalent to the conserving of special parity.
Then eq. (\ref{ik.10}) is represented as
\be
\lb{ik.10'}
\begin{array}{c}
\hat{K}_{54(16)32}(y,z; \overline{x}) \,
\hat{K}_{63(25)41}(x,z; y) \,
\hat{K}_{12(34)56}(x,y; \overline{z}) =
\\ \\
\hat{K}_{12(34)56}(x,y;z) \,
\hat{K}_{63(25)41}(x,z; \overline{y}) \,
\hat{K}_{54(16)32}(y,z;x)  \; ,
\end{array}
\ee
where we omit indices $(\tilde{x}, \, \tilde{y}, \, \tilde{z})$ and
introduce dressed reflection operators
$$
\hat{K}_{12(34)56}(x,y;z) =
R_{123}(x,y,z) \, \hat{K}_{34}(x,y) \,
R^{-1}_{456}(z, \overline{y}, \overline{x}) \; .
$$
Examination of relation (\ref{ik.10'}) leads to the conclusion
that 3d reflection equation can be written in the form of the
Yang-Baxter equation for dressed reflection operators.

It is tempting to investigate 3d reflection equation (\ref{ik.10})
which is independent of spectral parameters. Namely we have
\be
\lb{ik.11}
R^{-1}_{123} \, R_{541} \,
\hat{K}_{16}^{\tilde{y}\tilde{z}} \,
\hat{K}_{25}^{\tilde{x}\tilde{z}} \,
R^{-1}_{541} \, R_{123} \,
\hat{K}_{34}^{\tilde{x}\tilde{y}} =
\hat{K}_{34}^{\tilde{x}\tilde{y}} \,
R^{-1}_{456} \, R_{632} \,
\hat{K}_{25}^{\tilde{x}\tilde{z}} \,
\hat{K}_{16}^{\tilde{y}\tilde{z}} \,
R^{-1}_{632} \, R_{456} \; .
\ee
The simplest constant solution of this equation is
$\hat{K}_{12}^{\tilde{x}\tilde{y}} = (I_1 \otimes I_2) \otimes
(R^{\tilde{x}\tilde{y}})$, where $I_{1,2}$
are unit matrices and operators $R^{\tilde{x}\tilde{y}}$
satisfy the Yang-Baxter equation
$R^{\tilde{x}\tilde{y}} \, R^{\tilde{x}\tilde{z}} \,
R^{\tilde{y}\tilde{z}} = R^{\tilde{y}\tilde{z}} \,
R^{\tilde{x}\tilde{z}} \, R^{\tilde{x}\tilde{y}}$.
Note that one can rewrite (\ref{ik.11}) in the form
(cf. with (\ref{ik.10'}))
\be
\lb{ik.12}
\hat{K}_{54(16)32}^{\tilde{y}\tilde{z}} \,
\hat{K}_{63(25)41}^{\tilde{x}\tilde{z}} \,
\hat{K}_{12(34)56}^{\tilde{x}\tilde{y}} =
\hat{K}_{12(34)56}^{\tilde{x}\tilde{y}} \,
\hat{K}_{63(25)41}^{\tilde{x}\tilde{z}} \,
\hat{K}_{54(16)32}^{\tilde{y}\tilde{z}} \; .
\ee
where $\hat{K}_{12(34)56}^{\tilde{x}\tilde{y}} \equiv
R_{123} \, \hat{K}_{34}^{\tilde{x}\tilde{y}} \, R^{-1}_{456}$.
Moreover if we put
$$
{\mbox{\large $\sigma$}}_{12(34)56}^{\tilde{x}\tilde{y}}
= P_{12} \, P_{34} \, P_{56} \,
\hat{K}_{12(34)56}^{\tilde{x}\tilde{y}} =
P_{12} \, P_{34} \, P_{56} \,
\left( R_{123} \, \hat{K}_{34}^{\tilde{x}\tilde{y}} \, R^{-1}_{456}
\right) \; ,
$$
then we derive the relations:
\be
\lb{ik.13}
{\mbox{\large $\sigma$}}_{12(34)56}^{\tilde{y}\tilde{z}} \,
{\mbox{\large $\sigma$}}_{54(16)32}^{\tilde{x}\tilde{z}} \,
{\mbox{\large $\sigma$}}_{12(34)56}^{\tilde{x}\tilde{y}} =
{\mbox{\large $\sigma$}}_{54(16)32}^{\tilde{x}\tilde{y}} \,
{\mbox{\large $\sigma$}}_{12(34)56}^{\tilde{x}\tilde{z}} \,
{\mbox{\large $\sigma$}}_{54(16)32}^{\tilde{y}\tilde{z}} \; ,
\ee
which are similar to the defining relations for
the braid group.

\section{Conclusion}

There is a topological interpretation of the 2d
reflection equation as one of the defining
relation for the braid group in a solid handlebody ($R^3$ with an empty
tube) \cite{PPK}, \cite{KS}.
It is natural to expect that a topological interpretation
exists for the 3d reflection
equation as well, for the TE was connected with 2-knots in
4-dimensional space and 2-categories \cite{Khar}.

Another technical possibility refers
to the face models, when the Boltzmann`s
weight indices are situated inside volumes of the cubic lattice. There is
a transformation from the
vertex models to IRC (Interaction Round Cube) models using
intertwining vectors \cite{PPK,SBMS}
$\psi (\alpha ; a,\, b,\, c,\, d)$.
The corresponding reflection matrix (boundary Boltzmann`s weights) is
given by the relation
$$
K_{12} \, \psi (1\, ; d,\, e,\, f,\, c) \,
\psi (2\, ; f,\, g,\, b,\, c) =
$$
$$
=\sum_{h,\,j} \, Q(f,\,g|a,\,b,\,c,\,d,\,e|h,\,j) \,
\psi (2\, ; d,\, h,\, j,\, c) \,
\psi (1\, ; j,\, a,\, b,\, c) \,.
$$
The indices $a,\, b,\, c,\,...$ are inside corresponding volumes (cubes)
while the intertwiner structure similar to the $R$-matrix one.
The analogues of the relations (\ref{ik.1a}), (\ref{ik.1b})
for IRC models are
$$
A(g\bin{f}{b} a |z,y) \,
A(g\bin{b}{d} c |z,x) \,
A(g\bin{d}{f} e |y,x) =
$$
$$
= \sum_{h} \, W \left(g \, |\bin{ace}{dfb} | \, h; \, x,y,z \right) \,
A(h\bin{a}{c} b |y,x) \,
A(h\bin{e}{a} f |z,x) \,
A(h\bin{c}{e} d |z,y) \; ,
$$
$$
A(g\bin{f}{b} a |z,y) \,
A(g\bin{d}{f} e |x,y) \,
A(g\bin{b}{d} c |x,z) =
$$
$$
= \sum_{h} \, \overline{W}
\left(g \, |\bin{aec}{dbf} | \, h; \, x,y,z \right) \,
A(h\bin{e}{a} f |x,z) \,
A(h\bin{a}{c} b |x,y) \,
A(h\bin{c}{e} d |z,y) \; ,
$$
where $A(a\bin{d}{b} c |z,y) = \psi (1 \, ; a,\, b, \, c, \, d) \,
A_{1}(z,y)$. Using these formulas one can rewrite
(\ref{ik.4}) as tetrahedron reflection equation for IRC models.
One can write 3d reflection equation with the state variables
on the faces as well.

It is well known that the commutativity of the transfer matrices can be
seen for the non-periodic boundary condition using two reflection equations
(see \cite{KS}). Analogues procedure with peculiarities connected with TE
exists for 3d reflection equation too. In particular,
the important feature of the 3d RE (\ref{ik.4}) which we have to
apply here is the invariance of this equation
under the covariance transformation
$$
K_{12}(x,y) \longrightarrow
R^{-1}_{k1j}(y, \, u, \, \overline{x}) \,
R_{jn2}(x, \, y, \, u) \, K_{12}(x,y) \,
R^{-1}_{2ki}(u, \, \overline{y}, \, \overline{x}) \,
R_{i1n}(x, \, u, \, \overline{y}) \; ,
$$
where $u$ is arbitrary 2-vector and $i,j,k,n$ are indices of
auxiliary matrix spaces.

An algebraic foundation of the TE (as
quantum groups for the Yang-Baxter equations) is not
completely clarified up to
now (see however \cite{Kash}). We think that the new 3d
algebras presented in this paper will help in understanding of
the algebraic structures lying behind the tetrahedron and
3d reflection equations.

\vspace{0.5cm}

{\bf Acknowledgement.} \\
The authors thank I.G.Korepanov, P.N.Pyatov, O.V.Ogievetsky,
Yu.G.Stroganov and A.B.Zamolodchikov
for interesting comments on reflection
equations and especially S.M.Sergeev
for many explanations on the tetrahedron equation stuff. P.P.K. appreciate
useful discussions with L.Baulieu and friendly working conditions in
Laboratoire de Physique Theorique et Haute Energie as well as support
of C.N.R.S. and RFFI grant N 96-01-00851.



\begin{thebibliography}{99}

\bibitem{PPK} P.P.Kulish, {\em Yang- Baxter Equation and
Reflection Equations in Integrable Models, in Low-Dimensional
Models, in Statistical Physics and Quantum Field Theory},
Eds. H.Grosse and L.Pittner, {\it Lect. Notes Phys.}
{\bf 469} (1996) 125.
\bibitem{SMS} V.Mangazeev, S.Sergeev and Yu.Stroganov,
{\em The Tetrahedron Equation and Three-Dimensional
Integrable Models}, in Proc. of Workshop "Geometry
and Integrable Models", Eds. P.Pyatov and S.Solodukhin,
World Sci. (1996) p.3.
\bibitem{BB} V.V.Bazhanov and R.J.Baxter, {\em J. Stat. Phys.}
{\bf 69} (1992) 453; {\em ibid.} {\bf 71} (1993) 839.
\bibitem{Bax} R.J.Baxter, {\em Comm. Math. Phys.}
{\bf 88} (1983) 185.
\bibitem{Mai} J.M.Maillet, {\em Algebra i Analiz}
{\bf 6} (1994) 206.
\bibitem{Kor1} I.G.Korepanov, {\em Comm. Math. Phys.} {\bf 154}
(1993) 85.
\bibitem{H} J.Hietarinta, {\em J. Phys. A} {\bf 27} (1994) 5727.
\bibitem{LM} A.Liguori and M.Mintchev,
{\em J. Phys. A} {\bf 26} (1993) L887.
\bibitem{Zam} A.B.Zamolodchikov, {\em Zh. Eksp. Teor. Fiz.}
{\bf 79} (1980) 641 (English transl: {\em JETP} {\bf 52} (1980) 325);
{\em Comm. Math. Phys.} {\bf 79} (1981) 489.
\bibitem{BaSt} V.V.Bazhanov and Yu.G.Stroganov,
{\it Teor. Mat. Fiz.} {\bf 52} (1982) 105.
\bibitem{GZ} S.Ghoshal and A.B.Zamolodchikov,
{\it Int. J. Mod. Phys. A} {\bf 9}, No. 21 (1994) 3841.
\bibitem{FRT} L.D.Faddeev, N.Yu.Reshetikhin, and L.A.Takhtajan,
{\em Algebra i Analiz} {\bf 1} No.1 178 (1989);
English transl: {\em Leningr. Math. J.} {\bf 1} 193 (1990).
\bibitem{Kor} I.G.Korepanov, {\em Mod. Phys. Lett.} {\bf 3}
No.3 (1989) 201.
\bibitem{KS} P.P.Kulish, and E.K.Sklyanin, {\em J. Phys. A}
{\bf 25} 5963 (1992);
P.P.Kulish, and R.Sasaki, {\em Progr. Theor. Phys.}
{\bf 89} 741 (1993).
\bibitem{SBMS} S.Sergeev, H.Boss, V.Mangazeev and Yu.Stroganov,
{\em Mod. Phys. Lett.} {\bf A 11}
No.6 (1996) 491.
\bibitem{Khar} V.M.Kharlamov, {\em  Movements of straight
lines and the tetrahedron equations},
Preprint University of Pisa (1992).
\bibitem{Kash} R.M.Kashaev, {\em Algebra i Analiz},
{\bf 8} No.4 (1996) 63; \\
R.M. Kashaev and S.M.Sergeev, {\em On pentagon, ten-term,
and tetrahedron relations}, Preprint ENSLAPP - L-611/96 (1996),
q-alg/9607032.

\end{thebibliography}
\end{document}